\documentclass{llncs}
\usepackage{hyperref}
\usepackage{epsfig}
\usepackage{graphicx}

\usepackage{epsfig,alltt,tabularx,amsmath}
\usepackage{times}
\usepackage{hyperref}
\usepackage{law}
\usepackage{mystyle}
\usepackage{epsfig}
\usepackage{wrapfig}
\input epsf

{

\newenvironment{respar}{
  \begin{list}{}{\topsep 0pt}
     \item[]
}{
  \end{list}}
\def\In{\begin{respar}}
\def\Un{\end{respar}}
\fboxsep=10pt
\fboxrule=1pt

\newcounter{saved}

\def\save{\saveitem{saved}\addtocounter{saved}{-1}}

\def\resume{\stepcounter{saved}\setitem{saved}}
{\begin{rules} \resume}%
{\save \end{rules}}

\pagestyle{plain}
\pagenumbering{arabic}
\title{Dependable Management\\ of Untrusted Distributed
Systems}
\author{Naftaly Minsky}
\institute{Rutgers University, NJ, USA}

\begin{document}

\date{4/15/14}

\maketitle

\bibliographystyle{plain}
%\thispagestyle{empty}

%\tableofcontents

\begin{abstract}
 The conventional approach to  the online
 management of distributed systems---represented by such standards as SNMP for network management, and
 WSDM for systems based on \emph{service oriented computing} (SOC)---relies
 on the  components of the managed system to cooperate in the
management process, by providing the managers with the means to monitor their
state and activities, and to control their behavior.  
Unfortunately, the trust thus placed in the cooperation of the managed
components is unwarranted for many types of systems---such as systems based on SOA---making the conventional
management of such systems unreliable and insecure.

This paper introduces a radically new approach to the  management of
distributed systems, called 
  \emph{governance-based
management} (GBM), which is based on  a
 middleware that can govern the exchange of
messages between system components.  GBM has a substantial
ability to manage distributed systems, in a reliable and secure manner, even
without any trustworthy cooperation of the managed components. 
  And it can fully incorporate the conventional management techniques wherever
such cooperation can be trusted. GBM also supports a \emph{reflexive mode}
of management, which manages the management process itself, making it safer.
However, GBM is still a  work in progress, as it raises several open problems
that needs to be addressed before this management technique can be put to
practice. 

\end{abstract}

\vspace{-2.0ex}\s{Introduction}\label{ss-intro}
 For a complex, long-lived, distributed  system to be dependable
it must be managed continuously. That is,  the system needs to be
monitored on line, in order to detect inefficiencies, failures, and attacks; and
it must be controlled during system operation in order to deal with the most
critical of these problems. The importance of  this type of management---to
which we refer as \emph{reactive} (distinguishing  it from another mode of
management to be explored in this paper\footnote{Reactive management also
differs from general software management, which involves, testing, debugging,
people management, etc.}---has been well
recognized \cite{casati-soa-03,soa-roadmap}.

But the reactive management of distributed systems confronts a
\emph{fundamental impediment}. Namely, the would be managers\footnote{By
 ``managers'' we mean either people, such as operators, or software components
 designed for various management
tasks.}
 have little, if any, sway over the system components they need to  manage. 
That is, the actual behavior of these components---which may be dispersed all
over the Internet---is largely invisible to the managers, and cannot be
controlled  by them from afar.

 Most, if not all, existing mechanisms for the reactive  management of distributed
 systems attempt to bypass
this basic impediment
 by \emph{relying on  the components of the managed system to
  cooperate in the
management process}. Namely, the system components are trusted to provide
 managers with  means to monitor their state and activities, and to
control their behavior. 
 Standards have been developed to facilitate such cooperation by
 components.
The first of these standards was
SNMP (Simple Network Management Protocol) \cite{snmp90-11}, developed about 20
years ago for the management of networks.  
An analogous standard, called
   WSDM (Web Services
Distributed Management) \cite{ibm-wisdom}, has been devised more recently for
the management of systems based on \emph{service oriented computing} (SOC).
Moreover, the reliance
on the cooperation of components is common to most, if not all, recent approaches
 to  the reactive management of distributed systems. In particular,
  \emph{autonomic systems} \cite{whi04-1} have been conceived to
 be composed of \emph{``autonomic
 components''} that are to be designed in conformance with the
 policies of the system they are part of. 
Same is true for the following attempts: 
the concept of Self-Managed Systems by Kramer et al. \cite{kra07-1};
the various versions of  Policy Based Management (PBM) \cite{cho08-1,kum08-1};
and several attempts at management via   Computational   Reflection
\cite{cou08-1,and09-1}. Some of these works employ either SNMP
 or WSDM standards; other use different types of cooperation.

Unfortunately,  the trust thus placed by the conventional management
techniques in the cooperation of the managed components is
unwarranted for many types of systems.
This is the case, in particular, for the \emph{application layer}
 of many distributed systems whose components are highly heterogeneous, and
 evolving.  It
is also the case 
for the emerging loosely coupled and heterogeneous
distributed systems, whose component parts
 may run on different platforms, may be written in different languages, and may
 be designed, constructed, and even maintained under different administration
 domains. The concept of \emph{service oriented architecture} (SOA)---which is
prevalent in industry and in government, in particular for the support of
  virtual organizations (VOs) and grids---represents an outstanding
example of such systems.
 We  refer to  these kinds of  systems as having an open architecture, 
or just being \emph{open}\footnote{The term ``open
 system,'' as used here, has nothing to do 
 with the concept of open source.}, for short.
 In part, the term ``open'' reflect the fact that
 component of such a system may change dynamically, or leave the system,
 while new components may be added to it at any time.

There is little justification for trusting the components of these kinds of
systems to cooperate in  their management. 
Therefore, if the traditional management techniques, which relies on such
cooperation, is applied to  open system, \emph{it would not be dependable},
 and it would be \emph{insecure}, as we shall argue in \secRef{conventional}.
One clearly needs a very different approach to the management of such systems.

\textbf{The thesis of this paper} is that a significant part of system
management can be accomplished via suitable governance of the  exchange of messages between system
components---even if none of
these  components can be trusted to cooperate in the
management process. And that such  governance can be done scalabely,
and in a dependable and secure manner by means of an appropriate middleware.

This thesis leads us to the introduction of what we call \emph{governance-based
management} (GBM) mechanism, which does not depend on the cooperation of the
managed components, but is able to utilize such cooperation if it is deemed to
be trustworthy.  Moreover, the governance of message exchange, on which GBM is
based, enables it  to manage
 the management process itself, thus rendering this process safer---we refer to
 this important capability as  \emph{reflexive  management}.

GBM is thus strictly more general then the conventional, SNMP-like,
 management mechanism, as it can incorporates the latter. Also, GBM has a
 potential of being more reliable and more flexible than the conventional
 management mechanisms. But this paper is only the first step in the creation
 of GBM. It present the idea of GBM, which constitute a radical departure for
 the SNMP-like management, and its general architecture. 
But this idea  raises several problems, which are still open and require
further research, and it requires rigorous  experimental validation.

The rest of this paper is organized as follows.
\secRef{conventional} outlines the conventional approach to  system management,
and discusses its limitations.
\secRef{approach} describes the gist of our  approach to system
management.
 \secRef{s-RM}  discusses the properties  of the  middleware which are 
required
for it to be able to support GBM; and provides a brief introduction to
 the \emph{law-governed interaction}
(LGI) middleware, which satisfies  these properties, and which we 
employ for this purpose.
 \secRef{framework} introduces  a generic framework  for GBM.
\secRef{case} describes a simple case study, of 
 supermarket chain managed via GBM---it is a partial description of an actual
 implementation of such a system, and of its management via GBM, as a
 \emph{proof of concept} of this management mechanism.
\secRef{next} describes part of what  is yet to be done for the vision of GBM,
 to fulfill
its considerable promise.
 We conclude in \secRef{s-conclusion}.

\vspace{-2.0ex}\s{The Conventional Approach to  System Management,
and its Limitations}\label{conventional}
To be  more specific about the conventional approach to system  management,
   we   describe here broadly the  WSDM standard for the management of SOA-like
   systems.
Under this standard each managed component, or
\emph{service}\footnote{Although in SOA, the terms  ``components'' and
``service'' are often used interchangeably,
we will usually use the term ``component,'' for uniformity with other system
types.}, 
 is expected to  provide what we call a  \emph{management interface} (MI) (it
 is called  \emph{``management agent''} in WSDM).
 The MI is supposed to   provide managers with means for getting  information about
 the state and behavior of the component, and means for controlling it.
These means are called  \emph{managerial capabilities}, and they are classified
into  three
types: (1) \emph{properties} of the component, which managers may
examine; (b) \emph{events} that occur at a component, which the component itself
would communicate to the managers that subscribed to them; and (c)
 \emph{operations} that  managers may invoke by sending specified kinds of
 messages to a component, to control it.

A distinction is made  between \emph{component-specific} capabilities;
and \emph{common} capabilities that need to be provided uniformly by all
components, or by a substantial subset of them \cite{ibm-wisdom}.
  For example,
the component-specific capabilities provided by the MI of a printer
 may include
properties such as \TT{toner}, that represents  the current
toner capacity of the printer,   and \TT{printTime}
 that represents the average time it took this printer to report back that a
 print request has been served; and
operations such as
\TT{closeQueue} that blocks new printing requests from being 
accepted by the printer.  Common capabilities may include properties such as
\TT{CPU utilization} and \TT{inflow}, which represents the number of messages that arrived at any
given component during a specified window of time; and operations such
as \TT{remove}, which would remove the given component from the system.

We now identify two types of limitations of the conventional management: (1)
consequences of over-reliance on  the cooperation of the managed components;
and (2) the risk due to the power vested in the managers of a system.

\p{(1) The drawbacks of over-reliance on the cooperation of the managed components:}
 We maintain that conventional  management,  which relies on the
 cooperation of the managed components, tends not to be dependable and be
 insecure, when applied to systems whose components are not very trusted---like
 SOA-based systems, and like most systems at their application level.
Moreover, such management tends  to be inflexible even when applied to trusted systems.

\emph{(1.a) The lack of dependability:} When the code of a component is not
trustworthy, because its code is not known or because it evolves frequently
in unpredictable manner, then the  management
interface (MI) provided by this component is not likely to be trustworthy
either. 
Note that this is not a serious concern
when managing a relatively closed and stable system. This is also the case for
 network
management (subject to SNMP standard), because the vendors of hosts, routers,
and firewalls---the main manageable components at the network layer---can usually
be trusted to implement the required MIs.

\emph{(1.b) The lack of security:} It is  difficult to
protect the MIs of untrusted components against  malicious attacks.
This is particularly hard when
the components are  dispersed all over the Internet, and are maintained under
different administrative domains.
 Moreover,
one cannot ignore the possibility that the writers of certain components have
no interest in cooperating with the management process, and may thus provide
wrong capabilities intentionally. Consequently, some MIs may be corrupted and
would provide managers with wrong, possibly intentionally
misleading,  information. And it is obvious that management based on such
information can harm the system being managed.

\emph{(1.c) The lack of flexibility:} this problem has to do with the
\emph{common capabilities} that all MIs of a given system are expected to
provide. Any change in such capabilities, or any addition to them, needs to be
carried out by each of the heterogeneous system components. This is a very
laborious undertaking, and a very error prone one.

\p{(2) The risk of managerial power:}
 The ability of
managers to monitor and control the system under their care provides them with
 enormous power with respect to that system. If left unchecked,
such power can be easily abused by careless or malicious managers, or by a
lack of proper coordination between different managers. This is the case
whether the manager is a person---playing the role of sys-admin or of an
operator---or if it is a program, designed to carry out some managerial tasks
automatically, perhaps under autonomic management.
Indeed, the harmful effect that operator's errors
often have on the system they manage is very substantial,
as is well known by practitioners, and studied  systematic
in the context of web-services
 \cite{bia05-1}.

To mitigate such risks, one needs to manage the process of management itself. 
Such a \emph{reflexive management} should entail  things like: imposing restrictions on what
different  managers can see and do; imposing necessary constraints on the order
of operations that can be carried out by a single manager, or by a group of
them; and logging the activities of managers.
It should be pointed out that
 SNMPv3---the third version of the SNMP standard for
network management---addresses some of these issues via  conventional access
control. But such  a control is too 
 rudimentary for the task at hand, in particular because it is not stateful,
 and not proactive---while both of these feature are critical for imposing
 constraints on the order of operations and on coordination between managers.

\vspace{-2.0ex}\s{The Gist of our  Approach to System
Management, and its Rational}\label{approach} 
As stated in the Introduction, the thesis that underlies our approach to the 
 management of untrusted distributed systems 
is that a significant part of such management
 can be accomplished by governing the interaction---via 
 messages  exchange---between the components of a given system, without the need to 
 rely on the
 cooperation of the managed components themselves.
This thesis is based, in part, on the observation that many, if not most, of
the WSDM-like capabilities currently used for system management
are \emph{communication-based}---that is, they can be defined purely in terms
of message exchange.
 For instance, among the example capabilities
mentioned above, the printer's property \TT{printTime} and the
operation \TT{closeQueue} on a printer are of this type; and so are the common
property \TT{inflow} and the common operation \TT{remove}. In particular,
the \TT{remove} operation, applied to a given component, can be carried out by
blocking all messages sent to, and by, a component to be removed, thus
 effectively separating it from the system. Of course, not all managerial
capabilities are of this type. Some capabilities, such as 
as the  \TT{toner}
property of a printer, or the \TT{CPU utilization} property of a host,   are defined in term of the 
 internal state and behavior of a component. Such
 \emph{internal capabilities}, must be provided by the component themselves, if
 they are to be usable for management.

This thesis is bolstered by the fact that complex societal
systems---like states, cities, enterprises, and  vehicular
traffic---are managed mostly  by observing and governing the interaction among people, and
between them and various inanimate entities (like stopping at a red light).
And this management is largely independent of the internal thoughts and private
behavior of people, which is usually not available to the management of societal
systems.

Based on this thesis, we plan to introduce a mechanism called  \emph{governance-based
management} (GBM), which
 operates primarily by analyzing and
governing the flow of messages between the distributed components of a
system---using a  middleware, suitable for this task.
One can distinguish between three functions of GBM, which would be carried out
in a unified manner: 

\emph{(1) The communication based capabilities} would be created under  GBM via an
analysis of the flow of messages in the system.
This, we maintain, can be done reliably, scalabely, flexibly and securely.

\emph{(2) For internal capabilities}, GBM  would utilize the conventional management
interfaces,  if they are provided
 by individual components, and if they are deemed to be trustworthy.
 And
 although these two types of managerial capabilities are to be provided in
 different ways, they are complementary, and can be used by managers in a
similar manner: by sending appropriate messages to individual
 components.

 \emph{(3) Reflexive management}
would be accomplished  by regulating the communication of managers with the 
components being managed, and with each other.

\p{The Limited Objective of this Paper:} The purpose of this paper is to introduce a
  \emph{generic framework for
GBM}; a framework that provides \emph{necessary} conditions for effective
management of open systems. It should be pointed out, however, that these
are \emph{only} necessary conditions for management. Effective management  also
requires strategies about what to monitor and when, and how to respond to
system failures. Such strategies are highly application dependent, and are
beyond the scope of this work, which focuses on general issues of management of
distributed systems.

\vspace{-2.0ex}\s{On The Middleware Underlying  GBM}\label{s-RM}
Effective governance-based management requires a powerful middleware to
be based on, which can regulate the exchange of messages between the components
of an open system.
 In particular, this mechanism needs to
satisfy the following conditions: (a) it must be
stateful and proactive, to be able to represent  communication based
capabilities, and to regulate coordination between managers; (b) it must be
decentralized, for scalability; (c) it must itself be dependable and secure, for
obvious reasons; (d) it must support multiple policies, allowing
for smooth interoperation between them; and (e) it must provide for
 policies to be incrementally composed into what we call \emph{conformance
 hierarchies}. (The meaning of, and reason for, the last two requirements will become
 evident in due course.)

\begin{wrapfigure}{r}{3.5in}
\begin{center}
\begin{tabular}{c}
 \epsfig{figure=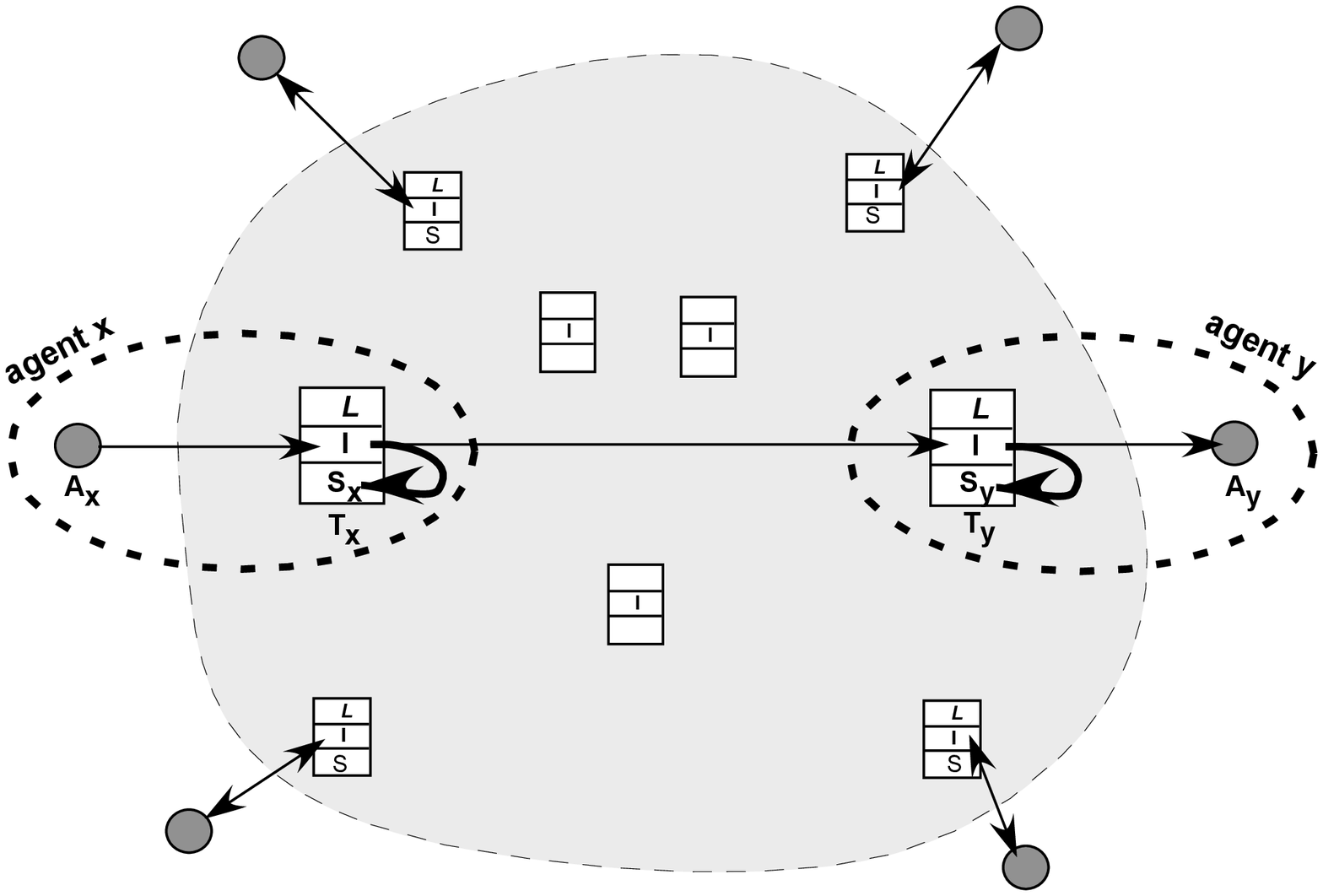,width=3.5in}
\end{tabular}
\vspace{-2mm}
\end{center}
\caption{Interaction via LGI: Actors are depicted by circles,
interacting across the Internet (lightly shaded cloud) via their private
controllers (boxes), one operating under law \EL, and the other under law \EL'.  Agents are depicted by dashed ovals that enclose
(actor, controller) pairs.
 Thin arrows represent messages, and thick arrows
represent modification of state.
\label{fig-overview}
\vspace{-3mm}}
\end{wrapfigure}

These  requirements are not easy to satisfy, and
 as has been demonstrated in
\cite{min12-2},
 they are  largely not satisfied by conventional
\emph{access control} (AC)---the currently dominant approach to governance of
distributed systems.  This is true for contemporary industrial AC
mechanisms such as  Tivoli and XACML \cite{god05-1}; for middlewares
such as CORBA  and J2EE; as well as for recent research 
mechanisms such as Oasis \cite{hay98-1}, SPL
\cite{rib07-1}, or BAM \cite{jen04-1}.
Even Ponder \cite{dam01-1}, perhaps the first attempt to support system
management via governance, is not suitable for our purpose. This is, in part,
 because Ponder
separates the support of management (via its concept of \emph{obligation}) from
its unstateful treatment of access control, and also because it does not
support conformance hierarchy, which turns out to be  essential for flexible management.

For this paper  we employ
a regulatory mechanism
 called
\emph{law-governed interaction} (LGI) \cite{min99-5}.
   This mechanism,  whose
prototype has been released  for public use, goes well beyond
conventional access control, and it satisfies all the above mentioned
requirements,  as has been demonstrated in \cite{min12-2}.
 The LGI mechanism is outlined below.

\vspace{-2.0ex}\ss{The Law-Governed Interaction (LGI) Mechanism---an
  Overview}
\label{ss-lgi}
Broadly speaking, LGI is a regulatory mechanism  that enables an
open and heterogeneous group of distributed \emph{actors} to engage in
a mode of interaction \emph{governed} by an explicitly specified and
strictly enforced policy, called the \emph{law} of this group. By
``actor'' we mean an arbitrary process, whose structure and behavior
is left unspecified; and an actor engaged in an LGI-regulated
interaction, under a law \CAL{L}, is called an \CAL{L}-$agent$.
LGI thus turns a set of disparate actors, which may not know or trust each
other, into a \emph{community} of agents that can rely on each other to comply
with the given law \CAL{L}.
 This is done via a distributed collection of  generic components called
\emph{private controllers}, one per \CAL{L}-agent, that
are trusted to  mediate all interaction between these agents, subject to the specified law
\CAL{L} (as illustrated in  Figure~\ref{fig-overview}).

All told, 
LGI goes well beyond conventional access control, in its ability to
 cope with the increasing size, openness, and heterogeneity of  distributed systems.
 It is, in particular,  inherently decentralized, and thus scalable
even for a wide range of stateful policies. And it is very general.  A prototype
of LGI has been recently released, and has a growing community
of users.

 This section provides only a very
brief overview of LGI, hopefully sufficient for understanding the gist of this
proposal.  For more information, the reader is referred to the LGI tutorial and
manual \cite{min05-8}, and to a host of published papers---but perhaps the most
approachable text is a recent, and yet unpublished, abstract model of LGI
\cite{min12-2}.

\p{Agents and their Private Controllers:}
 An \CAL{L}-agent $x$  is a pair
\(x = \langle A_x, T^{\mathcal{L}}_{x} \rangle\), where $A_x$ is an
\emph{actor}, and \Tla{L}{x} is
its \emph{private controller}, which mediates the interactions of $A_x$ with
other LGI-agents, subject to law
\CAL{L}.
Each controller  \Tla{L}{x} maintains the
 \emph{control state} (or,  ``cState'') of
agent $x$, which  is some function of the history of interaction of $x$
with other community members.  The nature of this function, and its effect on
the ability of $x$ to communicate, is largely defined by the law
\CAL{L}.  The concept of law is defined in the following section.
 The role of the controllers is illustrated in Figure~\ref{fig-overview}, which
 shows the passage of a message from an actor $A_x$ to $A_y$, as it is mediated
 by a pair of controllers, first by
\(T^{\mathcal{L}}_{x}\), and then by \(T^{\mathcal{L}}_{y}\)---both operating, in this case, under the same law, although interoperability between different laws is supported by LGI as well.
 One of the significant aspects of such mediation is that under LGI every
 message
exchange involves dual control: on the sides of both the sender of a message,
and of its receiver.  

\ignore{
\begin{figure}
\leavevmode
\epsfysize=3.5 in
\centerline{\epsffile{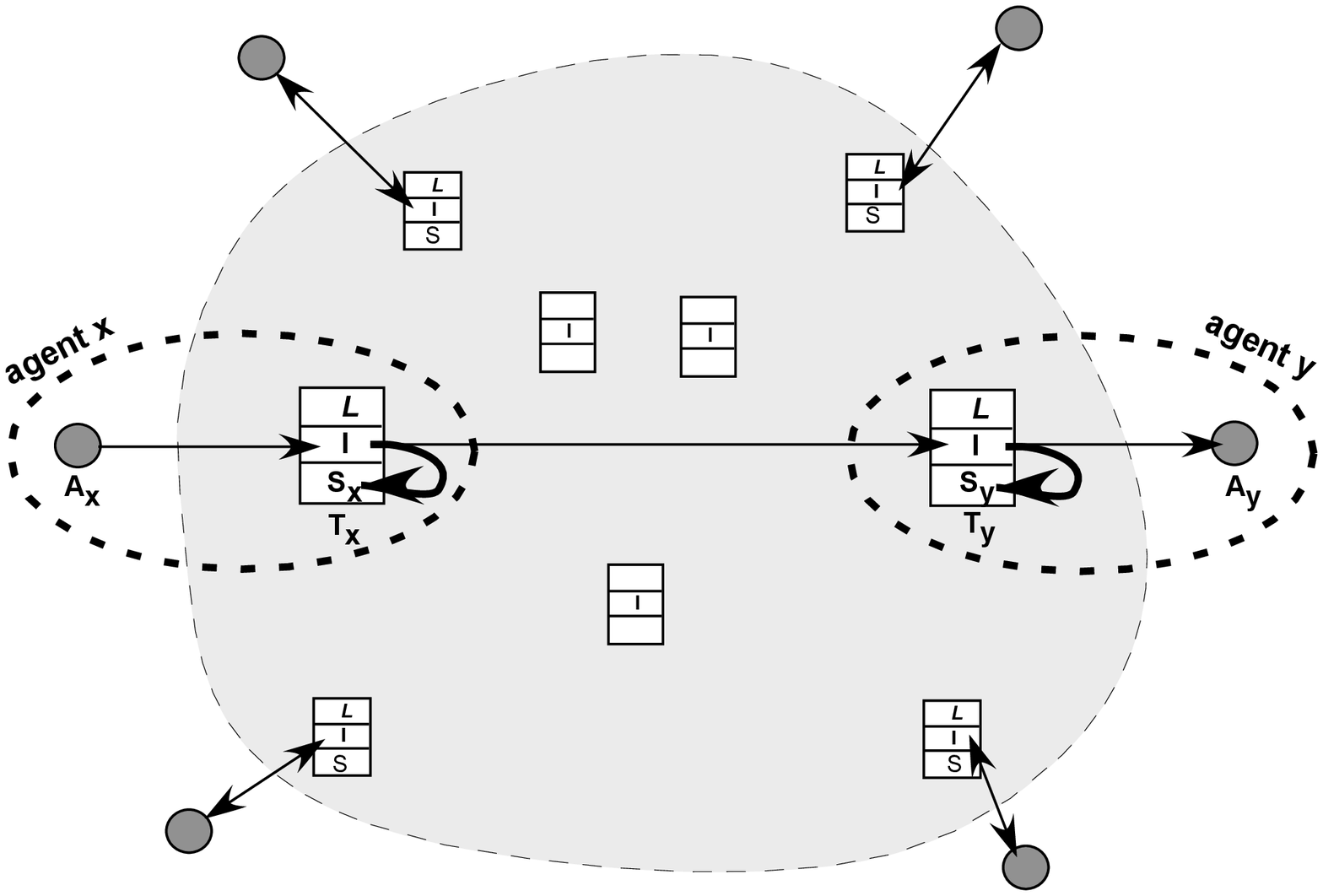}}
\caption{Interaction via LGI: Actors are depicted by circles,
interacting across the Internet (lightly shaded cloud) via their private controllers (boxes)
operating under law L.  Agents are depicted by dashed ovals that enclose
(actor, controller) pairs.
 Thin arrows represent messages, and thick arrows
represent modification of state.
\label{fig-overview}}
\end{figure}
}

It should be pointed out that private controllers
 are actually hosted  by what we call
\emph{controller pools}---each of which  is a process of computation that can
operate several (typically several hundreds) private controllers,  thus serving several different agents,
possibly subject to different laws. (Henceforth we will often refer to controller
pools as ``controllers,''  expecting the resulting  ambiguity to be resolved by the context.)
The set of controller-pools available to a given application (or a set of
application) is called a \emph{controller service} or CoS.
Interestingly, as we have shown in \cite{min99-5}, the use of duel controllers
actually reduces the overhead of mediation for communication over WAN---contrary
to what one could have expected.

\p{The Concept of Law Under LGI:}\label{s-laws}
An \emph{LGI law} (or, simply, a \emph{law}) is defined in terms of three
 elements: (a) a set $E$ of \emph{regulated events}; (b) a set $O$ of
 \emph{control operations}; and (c) the \emph{control-state} ($CS_x$)
 associated with each agent $x$.  More specifically, $E$ is the set of
 events---such as the sending and arrival of a message---that may occur at any
 agent, and whose disposition is subject to the law.
$O$  is the set of  operations that can
 be mandated by a law, to be carried out at a given agent, upon the occurrence
 of regulated events at it. In a sense, these operations constitute the
 \emph{repertoire} of the law---i.e., it is the set of operations that the law
 is able to mandate.
This set includes operations like forwarding a message, and updating the state
of a given agent. Finally, the 
 \emph{control-state}, or simply the state,  of an LGI agent is
the state maintained by the controller of this agent, which 
 is distinct from the internal state of its actor.
This state, which is initially empty, can change dynamically in response to the
various events that occur at it, subject to the law under which this agent
operates.

Now, The role of a law under LGI is to decide what should be done in response to
the occurrence of a regulated event at an agent operating under this law.
This decision, which is called the \emph{ruling of the law}, consists of a
 sequence of zero or more control operations from the set $O$.  More
formally, a law is defined as follows:

\begin{definition}[law]
Given a set $E$
of all  regulated events, a set $O$ of all control operations, and
a set $S$  of all possible
control-states, 
 a law \CAL{L} is a function:
$ \mathcal{L} : E \times S \rightarrow  O^*$
\label{def-law}
\end{definition}

In other words, \emph{a law maps every possible $(event,state)$ pair into a
sequence of zero or more control operations, which constitute the \emph{ruling}
of the law.}

Note that this definition  does not specify a language for
writing laws. This for several reasons: First,
because despite the pragmatic importance of choosing an appropriate
law-language, this choice has no impact on the semantics of the
model itself, as long as the chosen language is sufficiently powerful to
specify all possible functions of the form of Definition~1. Second,
by not specifying a law-language we provide the freedom to employ different
law-languages for different applications domains, possibly under the same
mechanism.  Indeed, the implemented Moses mechanism employs two different
law-languages, one based on the logic-programming language Prolog, and the
other based on Java.

\p{On the Basis for Trust Between Members of a Community:}\label{ss-trust}
For an \CAL{L}-agent $x$ to trust its interlocutor $y$  to
observe
  law \CAL{L}, it is
sufficient for $x$ to have the
 assurance that the following three conditions are satisfied: (a)
  the exchange  between $x$ and $y$ 
is mediated by correctly implemented private
 controllers \Ta{x} and \Ta{y}, respectively; (b) both  controllers 
operate under  law \CAL{L}; and (c) the
\CAL{L}-messages exchanged between $x$ and $y$ are transmitted  securely over
the Internet. 

The first of these conditions is the hardest to satisfy, and its support is one
of the main goal of this project. The other two condition are straightforward.
To ensure condition (b), that is that the interacting
 controllers  \Ta{x} and \Ta{y} operate under the same law,
 LGI adopts the following protocol:
When forwarding a message, a controller, say  $T_x$,  appends to it a
 \emph{one way hash}  \TT{H} of its law. The controller of the interlocutor, $T_y$ in
 this case, would accept this
as a valid \CAL{L}-message only if \TT{H} is identical to the hash of its own
law. Of course, such an exchange of hashes of the law can be trusted only if
 condition (a)  is satisfied.
Finally, to ensure the validity of condition (c), above, the messages sent
 across the Internet---between actors and their controllers, and between pairs
 of controllers---should be digitally signed and encrypted. These conventional,
 but rather expensive, measures have not been employed
in the current implementation of LGI. They are to be addressed under this
 project.

\p{The Local Nature of LGI Laws, and their Global Sway:}
Our concept of law differs structurally from the conventional concept of AC
policy, as discussed in \cite{min12-2}. One important characteristic of LGI
laws is that they are inherently
local.
Without going into technical details, locality means that an LGI law can be complied with, by each member of the community
subject to it, without having any direct information of the coincidental state
of other members.  This locality is a critical aspect of LGI for two major
reasons:
 First, because locality is necessary for decentralization of law
enforcement, and thus for scalability even for stateful policies. And second,
because locality facilitates interoperability between different laws, and
enables the construction of law-hierarchies, as has been shown in
\cite{min03-6}. 

 Remarkably, although locality constitutes a strict constraint on the structure
of LGI laws, it does not reduce their expressive power, as has been proved in
\cite{min05-8}. 
In particular, despite its \emph{structural locality}, an LGI law can have
\emph{global effect} over the entire \CAL{L}-community---mostly because all
members of that community are subject to the same law---and can, thus,  be used to
establish \emph{mandatory}, community wide, constraints.

\p{The Organization of Laws into \emph{Conformance Hierarchies}:}
LGI enables its laws to be organized into what we call 
 \emph{conformance hierarchies}.
Each such hierarchy, or tree, of laws $t($\CAL{L}$_0)$, is rooted in some law
\CAL{L}$_0$.  Each law in $t($\CAL{L}$_0)$ is said to be (transitively)
\emph{subordinate} to its parent, and (transitively) \emph{superior} to its
descendants.  And, given a pair of laws \CAL{N} and \CAL{M} in
$t($\CAL{L}$_0)$, we
 write \CAL{N}$\prec$\CAL{M} if \CAL{N} is subordinate to
 \CAL{M}. 
Semantically, the most important aspect of this hierarchy is that if
 \CAL{N}$\prec$\CAL{M} then \CAL{N} \emph{conforms} to \CAL{M}, in the sense
 that \emph{law \CAL{N} satisfies all the stipulations of its superior law
 \CAL{M}}.

This is a much more general concept of conformance than   adopted by some
 policy mechanisms (see \cite{dam01-1}, for example), where a policy $P$ is
considered in conformance with a policy $Q$, \emph{only} if $P$ is at least as restrictive
as $Q$.  
Briefly, the LGI's concept of 
 \emph{conformance hierarchy}  has two key properties: (a) it
 is  \emph{heterogeneous}, and (b) it is \emph{enforced}.
The hierarchy is heterogeneous with respect to conformance, in the following
sense: every law in the hierarchy
 can specify the degrees of freedom it
leaves to its subordinate (descendant); that is, each law   circumscribes the
degree and  manner in
which its descendant can deviate from it.
And the hierarchy is enforced by its very construction. That is, the very
definition of 
a law  \CAL{N} as subordinate to \CAL{M}, prevents  \CAL{N} from violating the
restriction imposed by  \CAL{M} on its subordinates.
The manner this is done has been defined
in \cite{min03-6}, and it  is too complex to describe here.

\p{Other  Features of LGI, and its Performance:}\label{sss-status}
We will list here some of the notable features of LGI, which we were not able
to discuss in this short overview, and will provide references to them for the
 interested reader.
These features are: 
(1) the concept of \emph{enforced obligation}, that provides LGI with important proactive
    capabilities; (2) the treatment of \emph{exceptions}, which provides LGI with
 fault tolerance capabilities; (3) the treatment of \emph{certificate},
    which is obviously necessary for the regulation of  distributed computing;
    and (4) \emph{interoperability}  between different laws. All these are discussed in
    the LGI Manual \cite{min05-8}.
Finally, we point out that the performance of LGI is discussed in
\cite{min05-8}.
In a nutshell, the overhead due to the LGI mediation is between 30 and 100
microseconds, for the types of laws we used in most of our studies.

\vspace{-2.0ex}\s{A generic Framework for  Governance-Based Management}\label{framework}
 A system managed under GBM (called GBMS)   is defined here as a
four-tuple $\langle B, M, T, LE\rangle,$ where $B$ is the \emph{base system}
being managed, also called the base layer (or B-layer) of the GBMS; $M$ is
the \emph{managing system}, also called the management layer (or M-layer) of
the GBMS; $LE$ is an ensemble of laws, organized into
a \emph{conformance hierarchy},
 that collectively enables the management of the system;
and $T$ is a set
of  LGI controllers, trusted to serve as the middleware underlying this mechanism.
It should be pointed out that this  framework assumes that the base system is 
 constructed from scratch to be managed under GBM. Applying GBM to
 legacy systems is still an open problem, to be addressed in the future (cf.,
 \secRef{next}. 

We now elaborate on this definition of GBMS by describing the following aspects of
it: (a) the general anatomy of a GBMS; (b) the structure of the hierarchical law-ensemble,
which is, in  a sense, the heart of the GBM framework;
(c) the deployment of a GBMS;  (d) its operation; and (e) its evolution.

\begin{figure}
\leavevmode
\epsfysize=3.2 in
\epsfxsize=5.9 in
\centerline{\epsffile{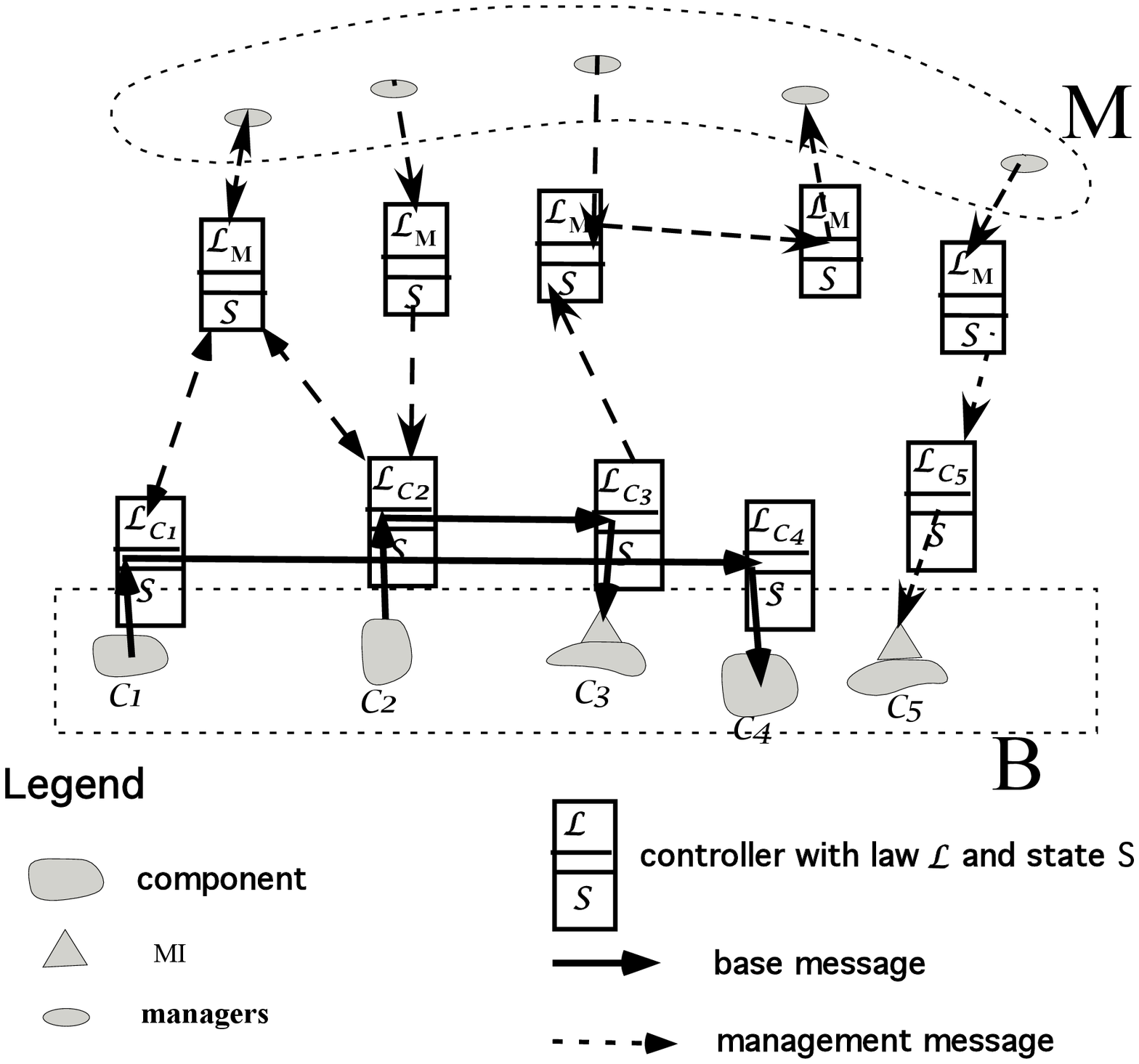}}
\caption{A Framework for Governance-Based Management
}
\label{fig-arch}
\end{figure}

\p{(a) The Anatomy of a GBMS:} 
The structure of a GBMS is depicted in \figRef{fig-arch}.  We
introduce here the various parts of this structure, and describe their roles in
the management process.
First, the active building blocks  of the base and management layers are autonomous entities we call \emph{actors},
which are treated essentially as black boxes under GBM.
  The actors $B$  are the   \emph{components} that comprise
  the system to be managed, and are  represented
in  \figRef{fig-arch} by
irregular shapes, which represents their presumed  heterogeneity.
 Some, but not necessarily all, such components would have a  WSDM-like
\emph{management interfaces} (MI) provided by
 them, which  are represented  in \figRef{fig-arch} by 
triangles at their top.
 The proposed mechanism would utilize such interfaces, when
they are available and considered trustworthy, but  it does not require that
all  components, or any of them, provide such interfaces. Note that the
internal capabilities of a component  $c$
that does not provide an MI, or whose MI is not considered trustworthy enough to be
used,   cannot be used for the management of $c$, but $c$
can still be managed via its communication-based capabilities defined for it by
the law ensemble.

The actors in $M$  are either  people (say
administrators or operators) that operate through software interfaces.; or they
are  software components that do
such things as log various events and analyze them. 
  This
framework  makes no  distinction between these two kind of  actors, referring
to both of them as   ``managers.''

 Every actor in either $B$ or $M$ is associated with  a
controller (depicted
by a box in  \figRef{fig-arch}), that operates under one of the laws
in the hierarchical law-ensemble $LE$.
Each such controller mediates the interaction of the actor it serves 
 with the rest of
the system, subject to the law under which this actor operates. And,  as we shall see below, the
controller serving  a component of $B$
 plays an analogous role to that of the traditional management interface
 (MI)
with respect  to what we have called communication-based
capabilities. Moreover, this controller mediates the interactions between
managers and the MI of of the, if any.

Finally, we  distinguish between two
types of messages: (1) base-messages, or \emph{b-messages} (depicted by solid lines in  \figRef{fig-arch}), are those exchanged between base
components via their corresponding
controllers; and (2) management messages, or \emph{m-messages} (depicted by  dashed lines in
\figRef{fig-arch}),
 are those exchanged between managers via their controllers, or between
 managers, via their controllers, and the controllers of B-components (generally not
 involving the  components themselves).

\p{(b) The Hierarchical Law Ensemble ($LE$) of a Managed System:}
 We introduce here an informal, and   rather schematic, description of the law
ensemble, which is based on   the concept of conformance
hierarchy briefly discussed in \secRef{ss-lgi}, and in more detail in
\cite{min03-6}. Some details about the possible content of 
such an ensemble are provided in \secRef{case}
in the context of a specific case study.

\begin{figure}
\leavevmode
\epsfysize=1.5 in
\epsfxsize=5.9 in
\centerline{\epsffile{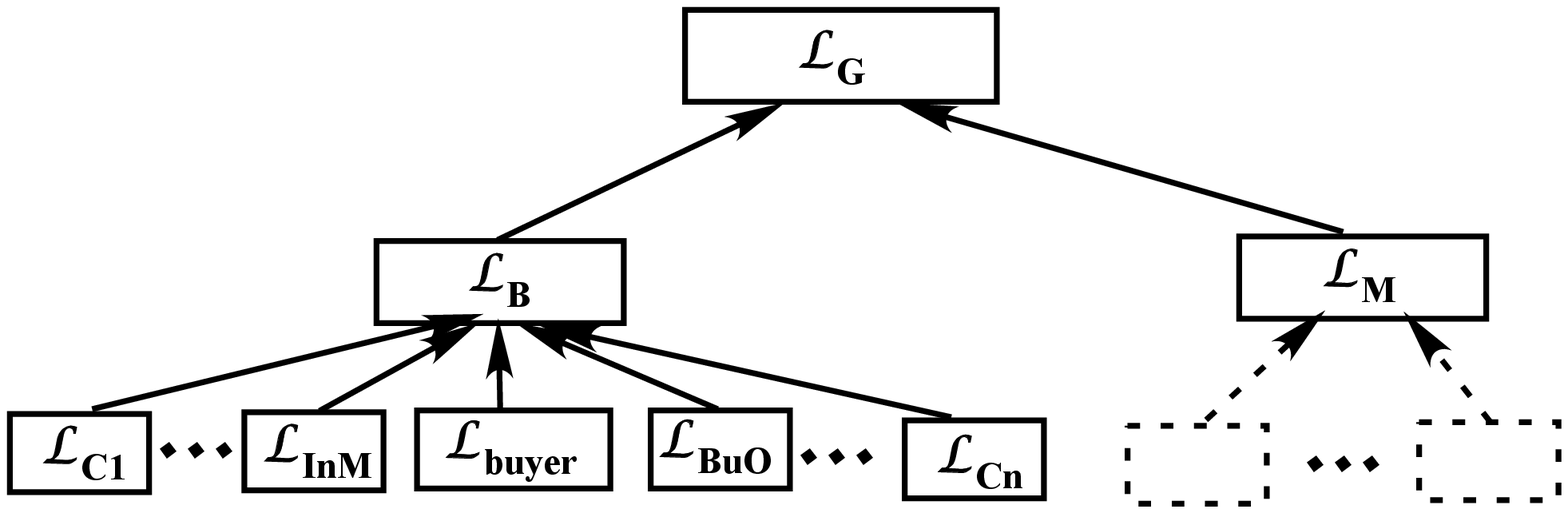}}
\caption{A Basic Hierarchical Law-Ensemble for Governance Based Management
\label{fig-laws}}
\end{figure}

The  law
ensemble of a GBMS is organized via the  concept of conformance
hierarchy of LGI introduced  formally and in details in 
\cite{min03-6}, and 
 described very briefly in \secRef{ss-lgi}, and utilized in various papers,
 such as \cite{min10-4}, for various application domains. The schematic structure of this ensemble 
is depicted in
\figRef{fig-laws}. The root of this hierarchy is
the law \law{G}, that governs the entire GBMS, because all other laws in this
ensemble are forced to  conform to this law
($G$ stands for ``global'').
Two laws in $LE$ are  directly  subordinate to \law{G}. They are:
  (a)  law \law{B} that govern the  B-layer of the  system, and (b)
law \law{M} that govern the M-layer. Subordinate
to law \law{B} there is a set of \emph{component-laws} \law{C_i}, one for each
component $C_i$ in $B$ (three of the components in this figure
represent specific B-components of the case study introduced in \secRef{case}).
Note, however, that several components that have the
same API, and require the same management capabilities, can operate under the
same law.

It is worth noting that the management 
 of complex systems, such as grids and virtual
organizations (VOs), which may span different administrative domains,
 is likely to require deeper law-hierarchies. In particular,
 law \law{M} may have several sub-laws (as suggested
by the dashed lines in \figRef{fig-laws}), one for each such domain. Similarly Law \law{B} may have
several sub-law, under which different divisions of the software system would
operate. But these are fairly straightforward generalizations of the simplified
architecture described above.

\p{(c) On the Deployment of a  System Managed under GBM:}
Recall that we are   assuming here that the base system is 
 constructed from scratch to be managed under GBM.
Under this assumption---which we plan to drop in in the future by considering
incremental deployment, and the
 management of legacy systems---the deployment of a  new GBMS starts with  the following
 steps, carried out sequentially: (a)  creation of a trustworthy controller service (CoS)
(or contracting the use of some public CoS, if there is one); (b)  definition of a root law \law{G} that would govern the entire
system;  and  (c) definition of  laws \law{B} and \law{M}, subordinate to \law{G},
that would govern the B-layer  and the M-layer, respectively.

Once these initial steps are carried out, one can add new components  to
the B-layer of the system, incrementally, and at any time.
 This is done, with an arbitrary component \law{C_i}, by first defining  an
appropriate law \law{C_i} for it,  subordinate
to \law{B}---such as the law \law{buyer} discussed in \secRef{case}; and then having component $C_i$ adopt a private controller  $T_i$
to mediate its interaction with the rest of the system under this law.
 As demonstrated  in our case study, a law can be designed such that only
 specified components can operate under it.

Finally,  it is worth pointing out that we have no means for forcing a component
of the base system to operate under any law subordinate to \law{G}, or indeed
to employ LGI for its communication. However, a component that does not satisfy
these conditions
may be barred from  communication with any component that does operate under a
law subordinate to \law{G}.

\p{(d) On the Operation of a GBMS:} The interactions between a pair of base components $C_i$ and $C_j$, is
 mediated by their respective controllers, subject to laws \law{C_i}
 and \law{C_j}, respectively. And each of these laws 
would take care of the specific  management capabilities defined by it, along
with the common capabilities defined by law \law{B} and \law{G}.

But at this point the reader may wander how can such \emph{interoperability}
between agent operating under different laws (or policies) be accomplished.
The conventional answer to this question---described
in \cite{gon96-1,bid98-1,mcd02-1,bon00-1}, in particular\footnote{Note
that these papers are using the term ``policy'' for what we call here
``laws''.}---is that one needs to form a \emph{composition}, say \law{C_{i,j}},
of these two laws, and mediate the interaction in question subject to such
composition. 
But, as shown in  \cite{mcd02-1},  such compositions tend to be computationally
hard,  even for simple types of policies. 
And one needs to create  a quadratic
number---in terms of the number of B-agents in a given system---of such
compositions. This is obviously impractical.

Fortunately, as we have shown in \cite{min03-6}, no such composition is
required for interoperation between laws in a hierarchical law-ensemble under
LGI.
This \emph{seamless interoperability} is due basically to two factors. The first  is the enforced nature of conformance
hierarchy, which provide the assurance to the two interlocutors,  that both
operate under the same superior laws, say \law{B} or even \law{G}.
If this common  heritage---which us recognizable under LGI---is
sufficient for two laws  to interoperate, no seams are needed between them.
The second factor is  dual control over
every message exchange (as explained in detail in  \cite{min03-6}).

\p{(e) On the Evolution of a GBMS:}
The GBM framework is very flexible with respect to several types of changes of
the system operating under it. In particular,  a change of the code of an existing component leaves the
    communication-based capabilities invariant, because they are determined by
    the law of this component, and not by its code.  (Note that, on the other
    hand,  a
    change in the code of a component can change its internal capabilities
    presented to the manager via its conventional MI; so our claim of flexibility pertains
    only to communication based capabilities.)

Second, the law ensemble is flexible with respect to the introduction of new
leaf laws that governs the interactive activities of individual components.
 Such changes are  basically local with respect to the law-ensemble,because it
 cannot effect other laws due to the conformance relation underlying the the
 hierarchical structure of the ensemble.

However, the GBM framework is not yet flexible with respect to certain kind of
changes of its law ensemble. We will take up this issue in \secRef{next}.

\vspace{-2.0ex}\s{Managing a Supermarket Chain: A Case Study}\label{case}
Consider  a distributed  enterprise system that serves a
 supermarket chain,  called Acme, which is comprised of several branches that
 include stores and the headquarter. And
 suppose that this software system consists
 of loosely-coupled,
heterogeneous, and 
  semi-autonomous components  (or services, if Acme is SOA-based).
To be a bit more specific suppose that Acme contains the following three types
of software components (one per branch) called: \TT{InM} (for ``Inventory
Monitor''),  \TT{buyer}, and  \TT{BuO} (for ``Budget Office'').
  An \TT{InM} monitors the
inventory level of the various products in its  branch, sending a purchase
request to its branch's \TT{buyer} for any product whose inventory is considered
too low.  The function of \TT{buyer} is to procure goods for its branch, in
response to the purchase request sent to it by \TT{InM}; this is done by sending
purchase orders (POs) to various vendors. But a \TT{buyer} is supposed to limit
its purchases by the budget assigned to it by messages it receives from the
budget office (\TT{buO}).  (We assume that the structure of the message exchanged
between these
 three components is pre-specified.)
 
Suppose now that the managerial objectives with respect to these components are
roughly the following: First, to measure the quality of service (QoS) provided by
the  \TT{buyer} components in all the branches. Second,
 to detect two kind of possible  \TT{buyer}'s misbehaviors,  and to respond
 appropriately to them. The  misbehaviors in question are:   (a) a \TT{buyer} 
issues POs for more money than allowed by its  budget; and (b)
 the buyer purchases products not requested by
  the \TT{InM} component of its branch. 
We will show now how these objectives, and others to be discussed below,
 can be addressed by an
appropriate law ensemble. But due to lack of space,
 the various laws  discussed below are described
very broadly, mostly in terms of  the provision made by them, without
spelling out the formal laws that establishes these provisions.
The  structure of our laws themselves will be illustrated only for a  very simple law
segment, which is spelled out in terms of  pseudo code that resembles the
Prolog-based law language of LGI.
Note also that only some of the following discussion deals with three
components named above, the rest of it deals with the entire Acme system, and is
fairly generic.

\p{The Root Law \law{G}:}
Being the root of the  hierarchical law ensemble  of the supermarket chain,
this law  governs the system globally, because  every law in LE
must conform to all its provisions, which are:

\noindent  \emph{(1) Uniform authentication and identification of Acme's actors:} For an actor to
      adopt a controller  under any given law \EL\ in LE---and thus be able to
      operate subject to a given law  \CAL{L} as part of Acme---it must certify itself via a certificate
      signed by a  CA specified by law \law{G}.  We assume that each such
      certificate  identifies  the \emph{name}, \emph{branch} and \emph{layer}
  of this actor; where the name is assumed to be unique for every branch, and 
 the layer identifies the actor as
 either a  $B$-agent (i.e., belonging to the base layer)  or an $M$-agent.
      This identifying triple would be stored in the state of the controller,
      and thus serve to identify the agent at run time. (Note that the GBM
      framework does not require digital authentications, but it supports it if it
is       considered necessary.)

\noindent \emph{(2) Sender identification:}  Every message sent by an actor, 
which is identified by the triple
       \emph{[name, branch, layer]}, would be prefixed by this triple. This
 is necessary  for the definition of various managerial capabilities
      at the controller of the receiver of a message, but it would be removed
      before the message is delivered to the target actor.

Note that this law provides  no managerial capabilities, but 
it facilitates the introduction of such capabilities by subordinate
laws.

\p{Law \law{B} of the Base Layer:}
This law has two complementary functions. First, to define a set of
\emph{common capabilities}, i.e., properties, operations, and events that would
be provided by the controllers of all B-agents, or a substantial subset of
them.
 Second, to define the  purview of the various managers.

\noindent
\emph{(i) Defining common capabilities:}
We first explain how a capability can be defined by this law.
Consider a property called \TT{POcount}, which represents
the number of  POs sent by a given agent.
This property can be defined by the 
  segment of law \law{B} displayed  in \figRef{law-count}.
 This  segment consists of two rules \ref{SC-1} and \ref{SC-2}, written here in pseudo code, 
 which is structurally
similar to the actual Prolog-based law-language of LGI.
The effect of \ruleRef{SC-1} is that any B-agent
 that sends a purchase order has the \TT{POcount} variable in the state of
its controller incremented by one, before the message itself is forwarded to
its destination. The effect of \ruleRef{SC-2} is that whenever a message of the form
 \TT{examine(``POcount'')}, sent by a manager, arrives at a B-agent, the
 current value of the \TT{POcount} variable would be forwarded to that
 manager. In other words, updated by rule   \ruleRef{SC-1}, to detect
 misbehaviors such as number of POs sent exceeds the number of purchase requests.

\begin{ruleset}{A  Segment of Law \law{B}---the  \TT{POcount} property, and
its examination\label{law-count}}
\begin{footnotesize}
\Rule  \textbf{UPON} sent(M) \textbf{IF}  M=PO(...)   \textbf{DO} [POcount<-POcount+1, forward]
\*
\label{SC-1}
\Rule  \textbf{UPON} the arrival of a message \TT{examine(``POcount'')} sent
by a manager \textbf{DO} [forward(POcount) to sender]
\* 
\label{SC-2}
\end{footnotesize}
\end{ruleset}

Other common properties can be defined in a similar ways, and common operations
and events  can be treated in an analogous manner.
For example, an operation
\TT{remove} can be implemented  as follows: (a)  B-messages sent or received by a given B-agent $x$ would be
blocked if the value of a variable \TT{blocked} in the controller of $x$ is 1; (b)
the variable \TT{blocked} is set to 1 upon the arrival of the message
\TT{invoke(remove)} sent by a manager.
Thus, a component is effectively removed from the system by this command, by
preventing it from communication with other system components.

\noindent \emph{(ii) Defining the purview of managers:} This law can be written to
 prevent a  B-component at a given branch $R$ from    accepting messages from
 managers not belonging to $R$.

\p{Law \law{buyer}:}
The function of this law, under which our  buyers are to operate,
 is to implement its own  \emph{specific managerial capabilities}.
 This is to be done in a similar manner to the implementation of common
capabilities by law \law{B}. 
For example, the following capabilities can be implemented by this law:

\noindent \emph{(1) A property \TT{budget}} that represents the balance of the purchasing budget
    of a given buyer $b$. It is to be computed by taking into account the
    budget-carrying messages sent to $b$ by the budget server; and the POs
    actually sent by the $buyer$. 

\noindent \emph{(2) A property \TT{avDelay}} that represents the average time difference between
    the arrival at $b$ of a purchase request, and the sending of the
    corresponding PO, or the sending of a rejection of the request, due, in particular, to lack
    of sufficient budget. This property could be used by managers to monitor the
    quality of service (QoS) provided by the  buyer.

\noindent \emph{(3) An event \TT{lawBudget}} that occurs at the buyer when its budget balance
    (i.e., the \TT{budget} property above) becomes smaller than a given
    threshold.

\noindent \emph{(4) access to conventional MI:} If a \TT{buyer} component has its conventional MI, which provide
some internal managerial capabilities with respect to this components, then
this law can enable the use of these capabilities by managers---as illustrated
for component \TT{C5} in \figRef{fig-arch}.

\p{Law \law{M} of Managers:} This law has two main functions. First, 
 to enable managers to use the capabilities
provided to them by the various B-agents. This is done by enabling agents
operating under this law to send to
B-agents at least the following three kinds of messages:
\TT{examine(property)}, \TT{invoke(operation)}, and
\TT{subscribe(event)}---where the parameters specify the managerial capability
being addressed. And  note that  by law \law{B}, of this example, B-component accept messages
only from managers at their own branch.

The second function of this law is to establish \emph{reflexive management}.
 In particular, law \law{M} can force specified
types of messages to be logged in a specified audit
trail.  It can also establish a way for providing different  managers with
different roles, and restrict what agents in each role can do. Moreover, this
law
can device protocols that managers have to follow when they interact with 
 each other, which can help
them  to coordinate their activities safely.

\vspace{-2.0ex}\s{What is Yet to be Done}\label{next} 
Although a prototype of GBM has been
implemented, and is being tested, there are some open problems to be solved and engineering
 work to be carried out, before GBM can be used for real applications. The
 following are outlines of some of the issues that need to be addressed, with
 some preliminary thought about how they can be approached.
The first subsection below discusses briefly some of the remaining open problems regarding
GBM; and the second subsection addresses the need to validate the efficacy of GBM.
\vspace{-2.0ex}\ss{Open Problems and Potential Extensions}

\p{Applying GBM to Legacy Systems}\label{legacy}
So far we have  assumed  that the base system is 
 constructed from scratch to be managed under GBM, requiring all
 base-components of the managed system to communicate via LGI, subject to the
 laws specified for them.
 To apply GBM to a legacy system, it has to be done in a
manner that does not require the code of the various components
 to be even aware
of the LGI middleware.  A similar task has been accomplished  \cite{ngu06-1} by
a group working on regulating
access to file systems via LGI,  within an
Intranets. This has been done  by intercepting communication by means of firewalls. 
 But such interception would not be trustworthy for a geographically  dispersed
 system over the Internet. So, one need a more general solution to this
 problem. 

We already did some  preliminary work  \cite{min09-5} on one approach to this
problem, which  requires a change in the LGI middleware, and has certain
drawbacks. 
 Another possible approach for
solving this important problem is based on the following observation\footnote{
I owe this idea to Dr. Josephine Micallef, from Telcordia Technologies.}:
 If we can apply GBM to systems based on Service Oriented
Architecture (SOA) principles, then we can apply it also to any
\emph{SOA-enabled} legacy systems---enabled through the use of facade
patterns that exposes the legacy system capabilities through SOA-based
interfaces.

\p{Bootstrapping the Management of the Controllers  Used for GBM:}\label{model-open}
It stands to reason that  the set $T$ of controller, which play a  central
 role in the GBM framework, needs to be managed as well. 
And it does not seem to make sense to   apply GBM for the management of its own
underlying structure.
 Fortunately there is no need to do so. Because the set of
controllers is stable and  completely uniform (even though they tend to operate under
different laws), and can thus be managed reliably as kind of network components
in a SNMP-like manner. 
To do that  one needs to build into the controllers of LGI appropriate
 managerial capabilities for
managing controllers---both reactively and reflexively.
So, the management of the base layer of a GBMS, and of its set of controllers
are to be done in different ways. But these two types of managements are
obviously  interrelated, and need to be integrated---and it is not quite clear
how this is to be done.

\p{Dealing with the Evolution of the Law Ensemble of an LGDS:}
As has been pointed out in \secRef{framework}, it is very easy to add, remove,
or change a leaf law of the hierarchical law ensemble---all these are
virtually local changes. But  changes of a non-leaf law, such as
\law{G} or  \law{B}, 
 are not easy to handle due to  their global effect.
There are, in fact, two different problems with such changes, discussed briefly
below.

One problem with the changing of non-leaf law, say \law{G}, is the effect that
it might have on the laws subordinate to it. Unfortunately, there is
 no automatic  way
for figuring out, in general, what this effect is, and what to do about it.
So, after changing law \law{G} one needs to reconsider manually each of its
subordinate laws. 
However, it is possible to identify  classes of  changes of a non-leaf law 
that has no effect on its subordinates, or where the consequence of a change
can be handled automatically. The challenge is to identify and 
 formally characterize such classes, and to develop tools for calculating
 the effect of change in non-leaf law on its subordinates.

The second problem with changing non-leaf laws exists in systems that must
operate continuously, and cannot be stopped when laws are being changed. In
such systems, a leaf law must be changed while the system operates---which we
call \emph{in vivo} evolution of the law (i.e., evolution in a living organism,
as it where). This is a particularly hard problem because of the distributed
nature of the LGI middleware. We have recently addressed and 
solved this problem for the
special case of a system operating under a single law \cite{min09-2}. The
challenge is to extend this solution for the evolution of a whole hierarchical
law ensemble.

\p{An Exploration of Managerial Techniques and  Patterns:}
To gain a better understanding of the potential inherent in GBM, with its
unified  support for  \emph{reactive},
\emph{reflexive} modes of management, it is necessary to explore various
managerial techniques and   styles.  Here we mention just one such technique,
that of that of reconfiguration.

Reconfiguration is a well known and important managerial technique.
It is also quite difficult to accomplish, particularly in a dynamic open
system. Indeed, a recent paper about reconfiguration by Zaras et. al
\cite{zar06-1} concluded that most current reconfiguration techniques are being
used ``in the context of stationary systems, where reconfiguration is centrally
controlled.''  They go on to develop a sophisticated reconfiguration technique
for a more dynamic and decentralized context. But, like most
current management techniques, this paper relies 
  on the cooperation of the base components in providing trustworthy 
  management interfaces (MIs), and is, therefore, not effective for open
  systems. Moreover, the Zaras paper does not  support reflexive mode of
  management, which is particularly important for reconfiguration tasks
  that require synchronized changes in many parts of the systems. Such tasks
  often require tight coordination between different managers.

\vspace{-2.0ex}\ss{Experimental Validation of the Efficacy of GBM:}
Although GBM has been applied experimentally to a toy system---the basis for 
the case study  in \secRef{case}---this is far from sufficient for the
validation of the efficacy of such a radically new  approach  to system management.
Such validation requires applying GBM to a real---or at least
realistic---distributed system, and compare this new mode of management to the
conventional management technique, such as under WSDM.
This requires the ability to apply GBM to legacy systems, which is work in
progress in our lab. And it cannot, practically speaking, be done at a
university. A plan is under way to conduct such an experiment in collaboration
with Telcordia Technologies, under the leadership of  Dr. Josephine Micallef.
But a single experiment of this kind is probably not sufficient for the validation
of GBM, and we hope that the publication of this paper will encourage large
scale experimentation by others.

\vspace{-2.0ex}\s{Conclusion}\label{conclusion}
We have introduced in this paper the concept of governance-based management
(GBM) for distributed systems.
 Besides being more
reliable and more flexible than conventional management techniques,
particularly when
applied to open systems, it also support a critical  new modes of system
management. Namely, beside the conventional reactive management, GBM supports 
\emph{reflexive} management, which controls the management process itself, making it
safer.  Furthermore, GBM can incorporate conventional management standards like
SNMP and WSDM wherever the compliance is deemed trustworthy.

Although an experimental prototype of the proposed GBM mechanism has been
implemented,  and tested as a proof of concept,
 there are some open problems to be solved, and comprehensive
testing to be carried out, before GBM can be used for real application. We have
 outlined some of the issues that need to be addressed,
 with some preliminary
 thought about possible approach to them.

\bibliography{../../../writing-tools/biblio,monitoringRefer1,costel-biblio,bills}
\end{document}